# Two-Dimensional Group-IV Chalcogenide $Si_2Te_2$ film: A New Quantum Spin Hall Insulator with Sizable Band Gap

Run-wu Zhang,[a] Wei-xiao Ji,[a] Chang-wen Zhang*,[a] Ping Li,[a] and Pei-ji Wang[a]

[a] School of Physics and Technology, University of Jinan, Jinan, Shandong, 250022, People's Republic of China

**Abstract:**

Quantum spin Hall (QSH) effect are promising for achieving dissipationless transport devices due to the robust gapless states inside insulating bulk gap. However, the current realized QSH insulators suffer from extremely ultrahigh vacuum or low temperature. Here, by using first-principles calculations, we discover group-IV chalcogenide $Si_2Te_2$ film to be a 2D QSH insulator with a fundamental band gap of 0.29 eV, which is tunable under external strain. This nontrivial topological phase stems from band inversion between the Si-$p_{x,y}$ and Te-$p_{x,y}$ orbitals, demonstrated by a single pair of topologically protected helical edge states with Dirac point locating in the bulk gap. Notably, the characteristic properties of edge states, such as the Fermi velocity and edge shape, can be tuned by edge modifications. Additionally, the $h$-BN semiconductor is an ideal substrate for experimental realization of 2D $Si_2Te_2$ film, without destroying its nontrivial topology. Our works open a new route for designing topological spintronics devices based on 2D silicon-based films.

* Correspondence and requests for materials should be addressed to: ss_zhangchw@ujn.edu.cn

Two-dimensional (2D) topological insulators (TIs), also known as quantum spin Hall (QSH) insulators, are a new class of quantum materials in which, even though the bulk structure is insulating, the edge supports spin-polarized gapless states with Dirac-like linear energy dispersion protected by time-reversal symmetry (TRS)[1-3]. The robustness of edge states against nonmagnetic impurities makes 2D TIs better suitable for coherent spin-transport related applications. The prototypical concept of QSH effect is first proposed by Kane and Mele in graphene [4, 5], in which the spin-orbital coupling (SOC) opens a band gap at the Dirac point. However, the associated gap due to rather weak SOC is tiny (~$10^{-3}$ meV), which makes the QSH state in graphene only appear at an unrealistically low temperature. Quantized conductance through QSH state has only been experimentally observed in HgTe/CdTe[6, 7] and InAs/GaSb[8, 9] quantum wells at the ultralow temperature. The search for new classes of 2D TIs with sizable band gaps has thus become a challenge of urgent importance.

Several 2D materials have been predicted to overcome the issue of small bulk band gap, including germanene[10], stanene[11], bismuth[12], 2D transition-metal dichalcogenides [13–15], III-V bilayers [16], BiF [17], $Bi_4Br_4$ [18], $ZrTe_5$ and $HfTe_5$ [19], and 2D transition-metal halide [20]. Especially, following the experimental synthesis of hydrogenated germanene (GeH) [21], a fully chemical adsorption on honeycomb structure, e.g., Bi/Sb/As [22–25], Ge/Sn/Pb [26–28], as well as III-V films, [29, 30] have been proposed to host QSH effect with enhanced band gaps. For example, the band gap of halogen or organic molecule decorated stanene can reach a gap much as large as 0.30 eV, [26-35] making those 2D materials good candidates for the realization of QSH effect at room temperature. In spite of this successful progress, the topological quantum states in 2D films are still challengeable experimentally.

2D MXenes, [36-38] which are a set of 2D layered transition-metal carbides and nitrides reported recently, characterizes with offering many potential applications in novel electronic devices. More interestingly, Weng *et al.*[39] find that the oxygen functionalized MXene can turn into a native 2D TI with a gap of 0.2 eV, opening a new avenue to design 2D topological phase. Motivated by this work, here we propose a new large-gap 2D QSH insulator in group-IV chalcogenide $Si_2Te_2$ film, its stability is confirmed by phonon spectrum. The band gap is found to reach 0.29 eV, which can be further tuned under external strain. The physical origin of nontrivial topological phase stems from band inversion between the Si-$p_{x,y}$ and Te-$p_{x,y}$ orbitals, which is demonstrated by a single pair of topologically protected helical edge states. Notably, the characteristic properties of edge states, such as the Fermi velocity and edge shape, can be tuned by edge modifications. Furthermore, the effects of substrates on topological properties are explored when it is grown on various substrates via the weak van der Waals interactions, like BN sheet. These findings may shed new light in future design of QSH insulators based on 2D-silicon honeycomb lattice in spintronics.

First-principles calculations are performed by using density functional theory (DFT) [40] methods as implemented in the Vienna ab initio simulation package (VASP) [41]. The projector-augmented-wave (PAW) potential [42, 43], Perdew-Burke-Ernzerhof (PBE) exchange-correlation functional [44], and the plane-wave basis with a kinetic energy cutoff of 500 eV are employed. The Brillouin zone is sampled by using a 11×11×1 Gamma-centered Monkhorst–Pack grid. The vacuum space is set to 20 Å to minimize artificial interactions between neighboring slabs. During the structural optimization of $Si_2Te_2$, all atomic positions and lattice parameters are fully relaxed, and the maximum force allowed on each atom is less than 0.02 eV/ Å. SOC is included by a second variational procedure on a fully self-consistent basis. The screened exchange hybrid density functional by Heyd-Scuseria-Ernzerhof (HSE06) [45] is adopted to further correct the electronic structure. The phonon calculations are carried out by using the density functional perturbation theory as implemented in the PHONOPY code [46] combined with the VASP.

As shown in Fig. 1(a), the $Si_2Te_2$ film shares the honeycomb crystal with the space group of space group $P_{3m1}$ (No.164), which contains the spacial inversion symmetry and $C_3$ rotation symmetry perpendicular to 2D surface. In one unit cell, four atoms bond with each other to form four atomic layers stacked in the order of Te-Si-Si-Te, as illustrated by side view of $Si_2Te_2$ in Fig. 1(b). The equilibrium lattice constant of $Si_2Te_2$ has $a = b = 3.99$ Å and four-layer thickness $d = 4.21$ Å, which are determined from the energy minimization procedure. Phonon dispersion curve plays a very important role in correctly clarifying the lattice thermal conductivity. To check the dynamical stability of this structure, we have performed phonon spectrums, as shown in Fig. 1(c). One can see that all branches have positive frequencies and no imaginary phonon modes in the Brillouin zone, which confirms that $Si_2Te_2$ is thermodynamically stable.

The calculated electronic structures of $Si_2Te_2$ film in the vicinity of the Fermi level are shown in Fig. 2, in which the blue and the red colors stand for the Si-$p_{x,y}$ and Te-$p_{x,y}$ bands with and without SOC, respectively. It can be seen that when the SOC effect is not included, $Si_2Te_2$ is a conventional semimetal with a zero density of states whose valence band maximum (VBM) and conduction band minimum (CBM) touch at Γ point with the parabolic dispersions. Further orbital-projected analysis for the composition of the electronic states near the Fermi level reveals that VBM arise mainly from Te-$p_{x,y}$ orbitals, while CBM are composed of Si-$p_{x,y}$ orbitals, as seen in Fig. 2(a). When the SOC effect is fully considered, a direct band gap of 0.28 eV is opened up at Γ point, along with an indirect-gap of 0.23 eV. Another prominent feature is that the band inversion occurs, *i.e.*, the components of VBM and CBM states are exchanged clearly, which are opposite to band-state alignment in non-SOC case, as illustrated in Fig. 2(b). Since the PBE method usually underestimates the band gap, we also calculate the correctional band gaps using hybrid functional (HSE06).[45] As shown in Fig. S1 in the supplementary information, the nontrivial gap of $Si_2Te_2$ film can be enlarged to 0.29 eV, which are comparable with the ones of functionalized buckled Bi (111) (0.2 eV),[17-18], $ZrTe_5$ (0.1 eV)[19], as well as functionalized stanene (0.3 eV)[29,34-35]. This large nontrivial band gap of $Si_2Te_2$ without chemical adsorption, or field effects, is very beneficial for the future experimental preparation and makes it highly adaptable in various application environments.

The topological states can be further confirmed by calculating topological invariant $Z_2$ after the band inversion. Here we employ a recently proposed equivalent method for $Z_2$ topological invariant based on U(2N) non-Abelian Berry connection,[51] which allows the identification of topological nature of a general band insulator without any of the gauge-fixing problems. The evolution of Wannier Center of Charges (WCCs) to calculate the $Z_2$ invariant can be expressed as:

$$Z_2 = P_\theta(T/2) - P_\theta(0)$$

which indicates the change of time-reversal polarization ($P_\theta$) between the 0 and $T/2$. The evolution of the WCC along $k_y$ corresponds to the phase factor, $\theta$, of the eigenvalues of the position operator, $\hat{X}$, projected into the occupied subspace. Then the WFs related with lattice vector $R$ can be written as:

$$|R,n\rangle = \frac{1}{2\pi}\int_{-\pi}^{\pi} dk\, e^{-ik(R-x)}|u_{nk}\rangle$$

Here, a WCC $\bar{x}_n$ can be defined as the mean value of $\langle 0n|\hat{X}|0n\rangle$, where the $\hat{X}$ is the position operator and $|0n\rangle$ is the state corresponding to a WF in the cell with $R = 0$. Then we can obtain

$$\bar{x}_n = \frac{i}{2\pi}\int_{-\pi}^{\pi} dk\, \langle u_{nk}|\partial_k|u_{nk}\rangle$$

Assuming that $\sum_\alpha \bar{x}_\alpha^S = \frac{1}{2\pi}\int_{BZ} A^S$ with $S = I$ or $II$, where summation in $\alpha$ represents the occupied states and $A$ is the Berry connection. So we have the format of $Z_2$ invariant:

$$Z_2 = \sum_\alpha [\bar{x}_\alpha^I(T/2) - \bar{x}_\alpha^{II}(T/2)] - \sum_\alpha [\bar{x}_\alpha^I(0) - \bar{x}_\alpha^{II}(0)]$$

The $Z_2$ invariant can be obtained by counting the even or odd number of crossings of any arbitrary horizontal reference line. In Fig. 5(d), we display the evolution lines of Wannier function centers (WFC) for $Si_2Te_2$ film. It can be seen that the WFC evolution curves cross any arbitrary reference lines odd times, thus yielding $Z_2 = 1$.

To better understand the physical origin of QSH effect, we next do an orbital analysis around the Fermi level for $Si_2Te_2$ film. Here the band inversion process can be well explained in Fig. 3, where the energy levels at $\Gamma$ point near the Fermi level are mainly composed of Te-$p_{xy}$ and Si-$p_{xy}$ orbitals. The chemical bonding between Si and Te atoms make them split into the bonding and antibonding states, i.e., Si-$p_{x,y}^\pm$ and Te-$p_{x,y}^\pm$, where the superscripts + and − represent the parities of corresponding states, respectively. In the absence of SOC, as shown in Fig. 3(a), the bands near the Fermi level are contributed by the Si-$p_{x,y}^+$ and Te-$p_{x,y}^-$, with the Si-$p_{x,y}^+$ being above Te-$p_{x,y}^-$. When switching on SOC, the SOC effect makes the $p_{x,y}^+$ orbitals split into $p^+_{x+iy,\uparrow}$ & $p^+_{x-iy,\downarrow}$ and $p^+_{x-iy,\uparrow}$ & $p^+_{x+iy,\downarrow}$, while lowers Si-$p^+_{x-iy,\uparrow}$ & -$p^+_{x+iy,\downarrow}$ and raises Te-$p^+_{x+iy,\uparrow}$ & -$p^+_{x-iy,\downarrow}$. Due to the strong SOC strength of Si$_2$Te$_2$ film, the Si-$p_{x,y}^+$ and Te-$p_{x,y}^-$ band inversion occurs, resulting in a larger nontrivial band gap with $Z_2 = 1$. Noticeably, this mechanism is different from ZrTe$_5$/HfTe$_5$[19], 1T-MX$_2$, and 1T'-MoX$_2$ [13-15], where the SOC does not change band order, instead of only opening a sizable gap at the Fermi level.

Experimentally, the 2D films should be lie or grown on a substrate. It is reasonable to question how interfacial strain will affect the electronic and topological properties of Si$_2$Te$_2$ film. Figure 4(a) shows the evolution of direct gap ($E_\Gamma$) at Γ point and whole indirect band-gap ($E_g$) with respect to external strain. Here, we employ the biaxial strain on Si$_2$Te$_2$ maintaining the crystal symmetry by changing its lattices as $\varepsilon = (a-a_0)/a_0$, where $a$ ($a_0$) is the strained (equilibrium) lattice constants. One can see that the nontrivial band gap of Si$_2$Te$_2$ film is sensitive to the lattice variation, and the nontrivial band topology is preserved under strains of -6 – 2.3 %. Interestingly, a larger gap of 0.37 eV is obtained for 0.9 % strain. With increasing the lattice constant, $E_\Gamma$ monotonically decreases and reaches minimum value at a critical lattice ($\varepsilon_c = 2.3$ %). Beyond $\varepsilon_c$, the SOC interaction between Si and Te atoms becomes weaker significantly, decreasing the orbital hybridization between the bonding and antibonding states, and thus forms a TI-NI phase transition, as shown in Fig. 3(b). Similar trends are also found for HSE results, as shown in Fig. 4(b) and Fig. S2 in supplementary information. Here, we must point out that, although the bulk gap is enhanced for HSE method, the Te-$p_{x,y}$ and Si-$p_{x,y}$ band order is not changed, indicating that the nontrivial topology is robust to calculated method. As a consequence, the predicted band topology is qualitatively reasonable and reliable.

The most important character of the QSH insulator is helical edge states with spin-polarization protected by time-reversal symmetry (TRS). Thus, we calculate the topological edge states of Si$_2$Te$_2$ film by the Wannier90 package.[47] By using the maximally localized Wannier functions (MLWFs) and fitting a tight-binding Hamiltonian with these functions, the edge Green's function[48] of semi-infinite Si$_2$Te$_2$ is displayed in Fig. 5(a). One can explicitly see that each edge has a single pair of helical edge states in the bulk band gap and cross linearly at the *X* point. Besides, by identifying the spin-up (↑) and spin-down (↓) contributions in edge spectral function, the counter-propagating edge states exhibit opposite spin-polarization, making them compatible with the spin-momentum locking of 1D helical electrons. Remarkably, a sizeable bulk gap can stabilize the edge states against the interference of the thermally activated carriers, which is beneficial for observing room-temperature QSH effect. In addition, we find that the characteristic properties of edge state in QSH phase, such as carrier mobility, can be tunable by chemical edge modification, [49, 50] which is important to QSH application in spintronic device. To demonstrate these features explicitly, we construct a zigzag-type nanoribbon in Fig. 5(b), whose edge Te atoms are passivated by hydrogen (H) atoms to eliminate the angling bonds. The width of this nanoribbon is large enough to avoid interactions between the edge states of the two sides. Fig. 5(c) displays the corresponding band structures of $Si_2Te_2$ nanoribbon. In comparison to Fig. 5(a), the topological quantum states are drastically modified in two important ways: (*i*) The original extended gapless edge states become localized at Brillouin zone center in the *k*-space, similar to the edge states in HgTe/CdTe quantum well. [6, 7] Consequently, there is only one crossing point between the edge state and Fermi level within the energy window of bulk band gap, further indicating a band topology. (*ii*) The Fermi velocity of Dirac states, as reflected by the band slope, can be enhanced significantly with chemical edge modification. The Fermi velocity at Dirac point can reach ~ $1.2 \times 10^5$ m/s, comparable to that of $5.5 \times 10^5$ m/s in HgTe/CdTe quantum well, [6, 7] and larger than that of $3.0 \times 10^4$ m/s in InAs/GaSb quantum well, [8,9] which is helpful to practical applications in spintronics.

Finally, we construct a $Si_2Te_2$/BN heterostructure (HTS) to observe the QSH effect, since the BN sheet is an ideal substrate to assemble 2D stacked nanodevices in experiments.[52, 53]. Fig. 6(a) shows the geometrical structures of (2×2) $Si_2Te_2$ on (3×3) BN sheet, where the lattice mismatch is only about 1.32%. After full relaxation with van der Waals (vdW) forces[54], the $Si_2Te_2$ film retains the original structure with a distance of 3.47 Å for BN sheet. The calculated binding energy of this HTS is -98 meV, showing that it is typical vdW structure. In Figure 6(b) we present the calculated band structure of $Si_2Te_2$/BN HTS with considering the SOC effect. As expected, in these weakly coupled systems, $Si_2Te_2$ on BN sheet remains semiconducting, there is essentially no charge transfer between the adjacent layers, and the states around the Fermi level are dominantly contributed by $Si_2Te_2$ film. Interestingly, the band inversion is preserved here, maintaining the nontrivially topological feature.

In conclusion, we report a new large-gap 2D QSH insulator in group-IV chalcogenide $Si_2Te_2$ film, its stability is confirmed by phonon spectrum. The nontrivial band gap of $Si_2Te_2$ is 0.29 eV, large enough for room temperature applications, and it is tunable by moderate strains. The band topology stems from band inversion between Si-$p_{x,y}$ and Te-$p_{x,y}$ orbitals. A single pair of topologically protected helical edge states is established for $Si_2Te_2$ with the Dirac point locating in the bulk gap, and the odd numbers of crossings between edge states and Fermi level prove the nontrivial nature of $Si_2Te_2$ film. Notably, the characteristic properties of edge states, such as the Fermi velocity and edge shape, can be tuned by edge modifications. Furthermore, the effects of substrates on topological properties are identified when it is grown on various substrates via the weak vdW interactions, like BN sheet. These findings provide a new platform to design large-gap QSH insulator based on 2D layered films, which shows potential applications in spintronics devices.

________________________

**Acknowledgments:** This work was supported by the National Natural Science Foundation of China (Grant No.11434006, No.11274143, 61571210, and 11304121).

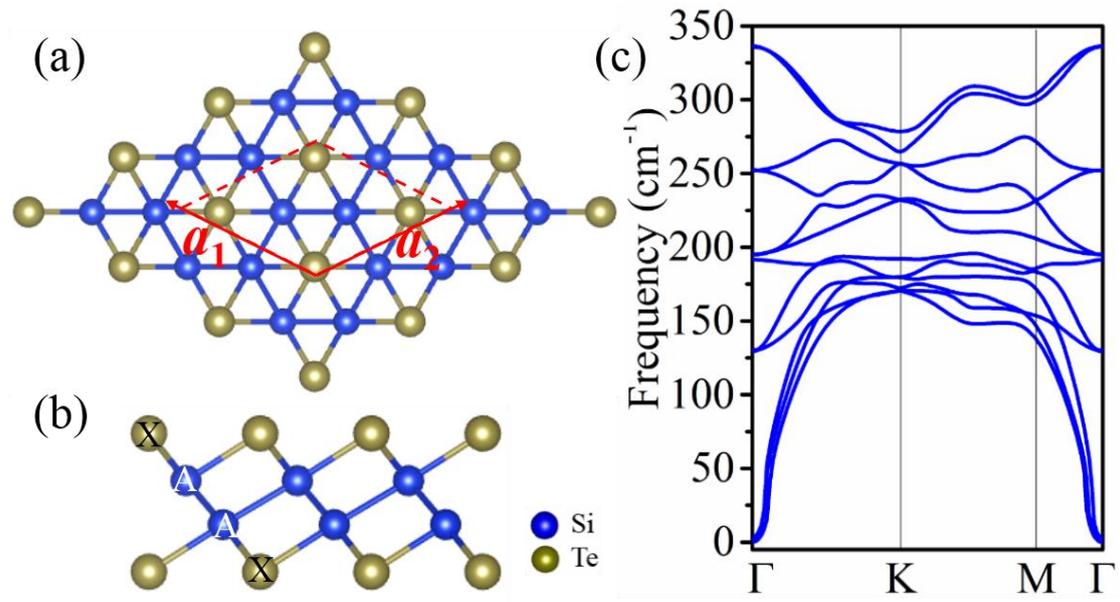

**Figure 1** (a) Side and (b) Top views of the atomic structure of Si$_2$Te$_2$. Dark yellow and blue balls denote Te and Si atoms, respectively. Shadow area in (a) presents the unit cell. Phonon band dispersions of (c) denote Si$_2$Te$_2$ film.

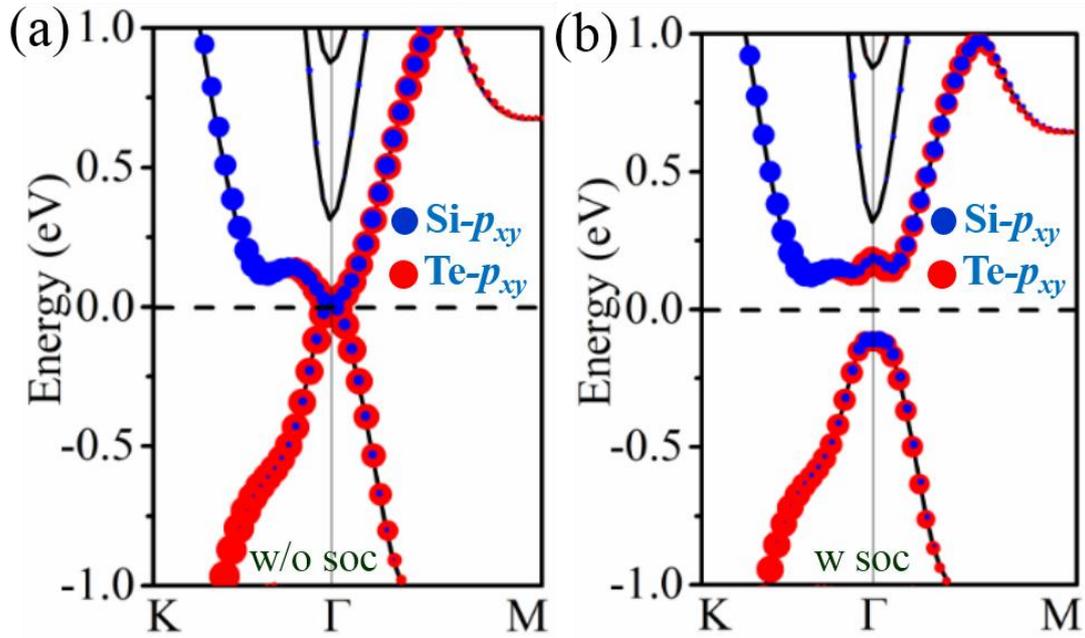

**Figure 2** Orbital-resolved band structures without SOC and with SOC of $Si_2Te_2$ (a) and (b) under equilibrium state, respectively. The red dots represent the contributions from the Te-$p_{x,y}$ orbitals and the blue dots represent contributions from the $p_{x,y}$ orbitals of Si atom.

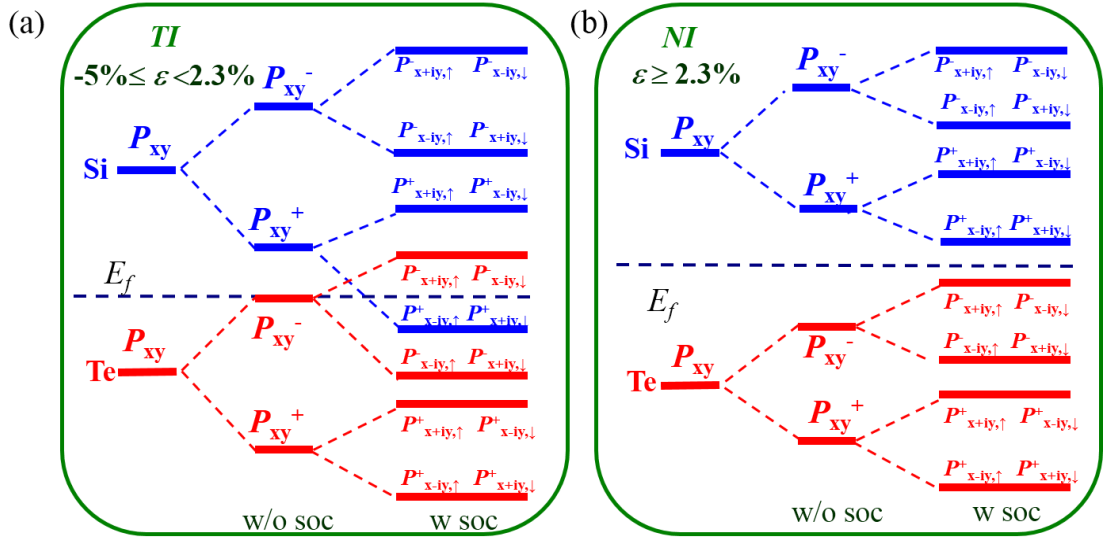

**Figure 3** The evolution of atomic Si-$p_{x,y}$ and Te-$p_{x,y}$ orbitals of (a) TI state and (b) NI state into the band edges at $\Gamma$ point is described as the crystal field splitting and SOC effect are switched on in sequence. The horizontal blue dashed lines in (a) and (b) indicate the Fermi level.

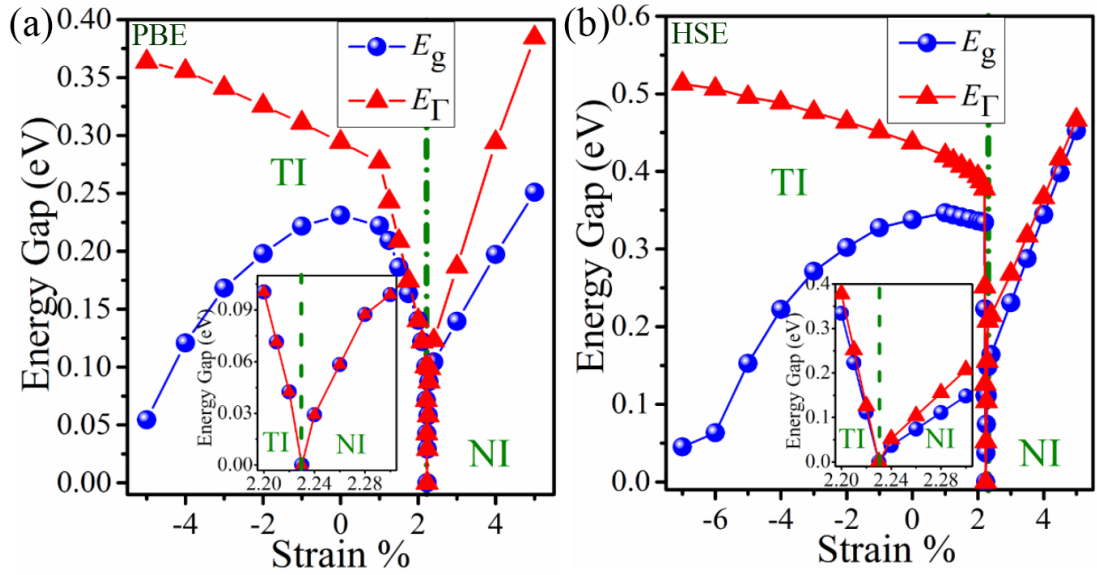

**Figure 4** The calculated band gaps at Γ point ($E_Γ$) and the global band gap ($E_g$) of Si$_2$Te$_2$ (a) with SOC as a function of external strain by PBE method. The band gaps at Γ point ($E_Γ$) and the global gap ($E_g$) of Si$_2$Te$_2$ (b) with SOC as a function of external strain by HSE method.

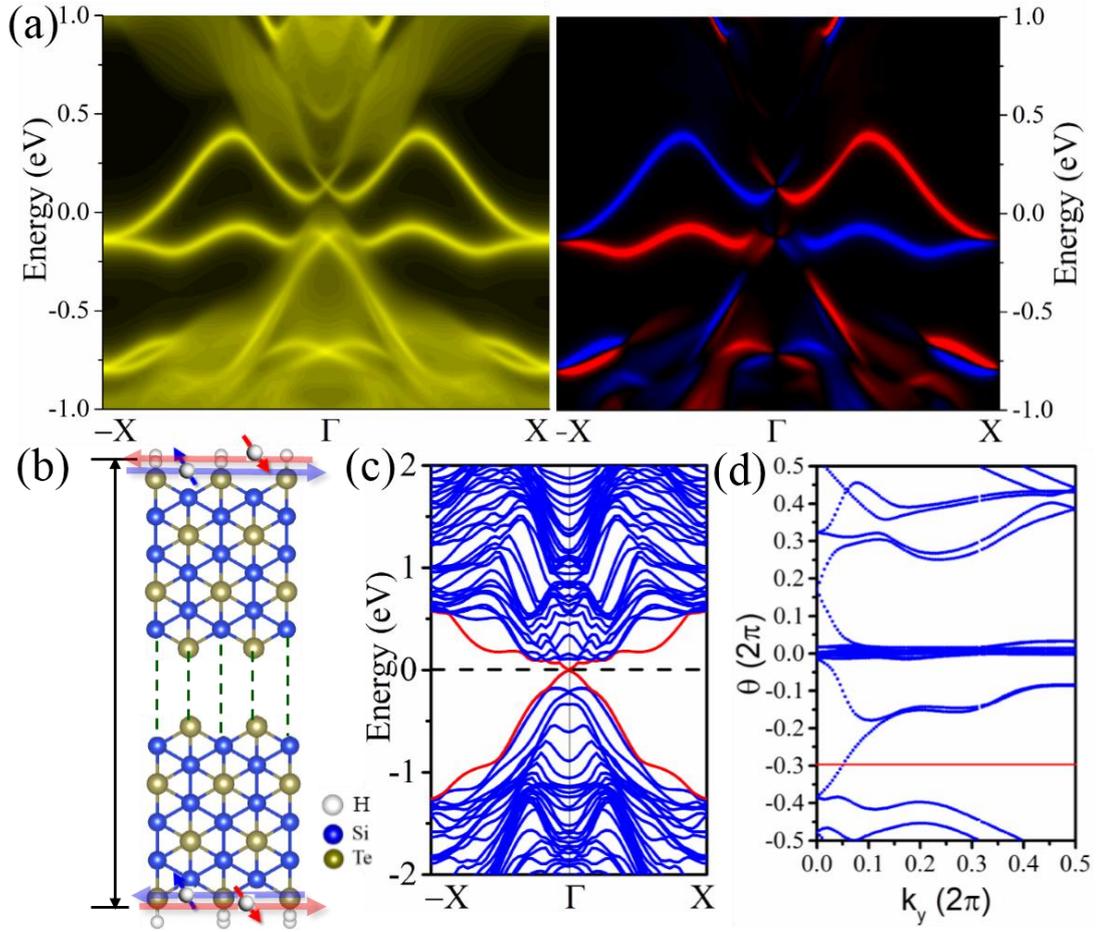

**Figure 5** Total (left panel) and spin (right panel) edge density of states for (a) $Si_2Te_2$. In the spin edge plot, red/blue lines denote the spin up/down polarization. Schematic atomic structure of $Si_2Te_2$ is (b); electronic band structures of the zigzag-type nanoribbons of (c) $Si_2Te_2$ with SOC. Evolutions of Wannier centers along $k_y$ are presented in (d). The evolution lines (blue dot lines) cross the arbitrary reference line (red dash line parallel to $k_y$) with an odd number of times, thus yielding $Z_2 = 1$.

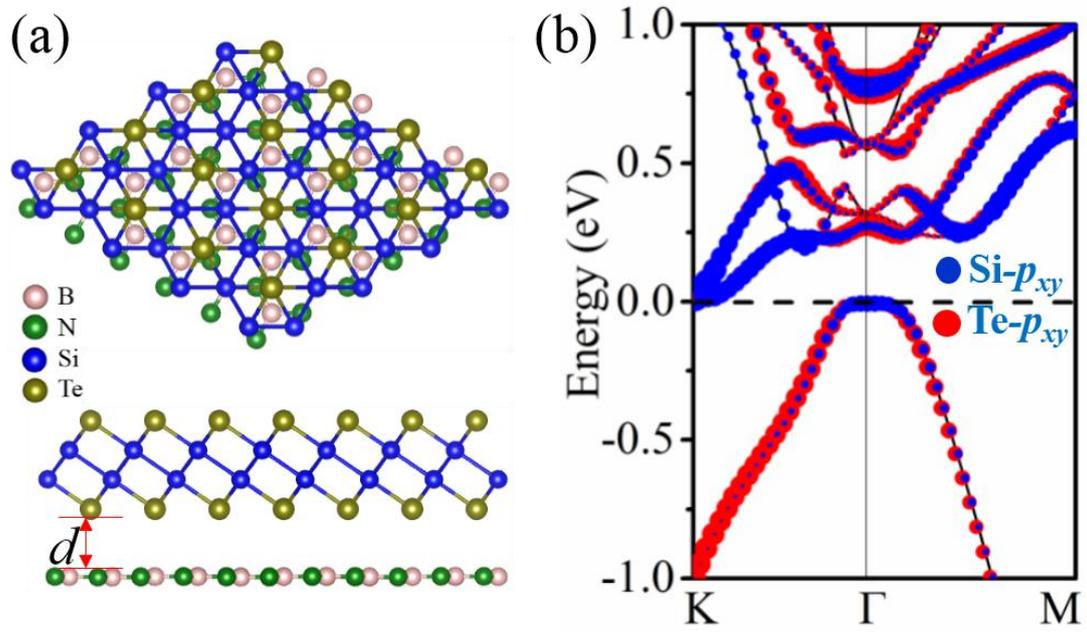

**Figure 6** (a) Crystal structure of Si$_2$Te$_2$ grown on BN substrate from the top and side views. (b) orbital-resolved band structure with SOC for Si$_2$Te$_2$/BN HTS.

# Supplementary Information for:

# Two-Dimensional Group-IV Chalcogenide Si$_2$Te$_2$ film: A New Quantum Spin Hall Insulator with Sizable Band Gap

Run-wu Zhang, [a] Wei-xiao Ji, [a] Chang-wen Zhang*, [a] Ping Li, [a] and Pei-ji Wang [a]

[a] School of Physics and Technology, University of Jinan, Jinan, Shandong, 250022, People's Republic of China

\* Correspondence and requests for materials should be addressed to: ss_zhangchw@ujn.edu.cn

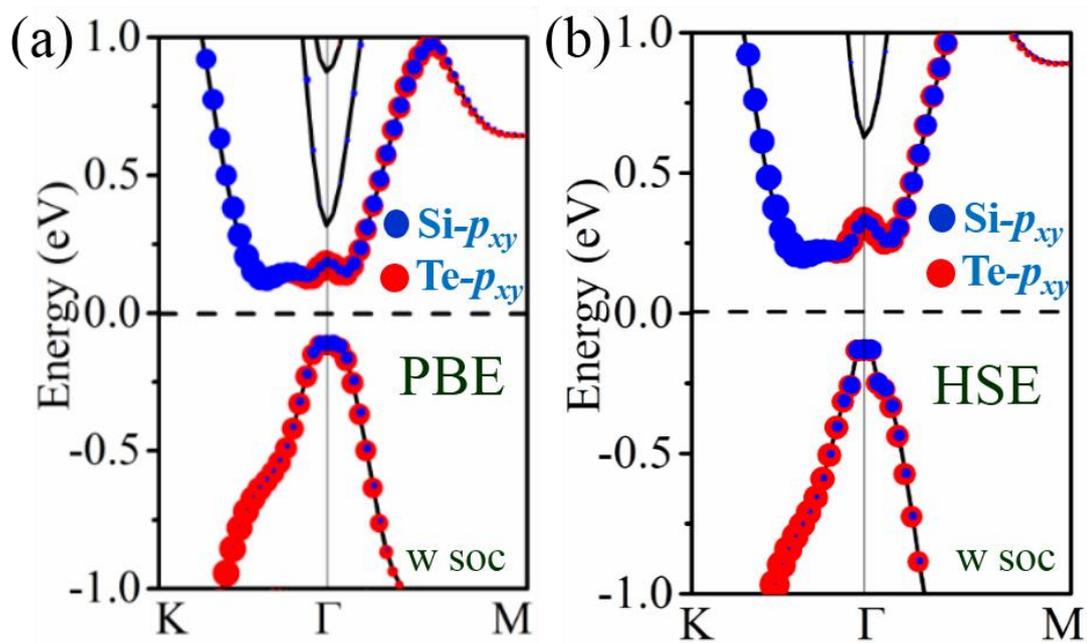

**Fig. S1**. Orbital-resolved band structures with SOC of $Si_2Te_2$ (a) PBE method and (b) HSE method, respectively. The red dots represent the contributions from the Te-$p_{x,y}$ orbitals and the blue dots represent contributions from the $p_{x,y}$ orbitals of Si atom.

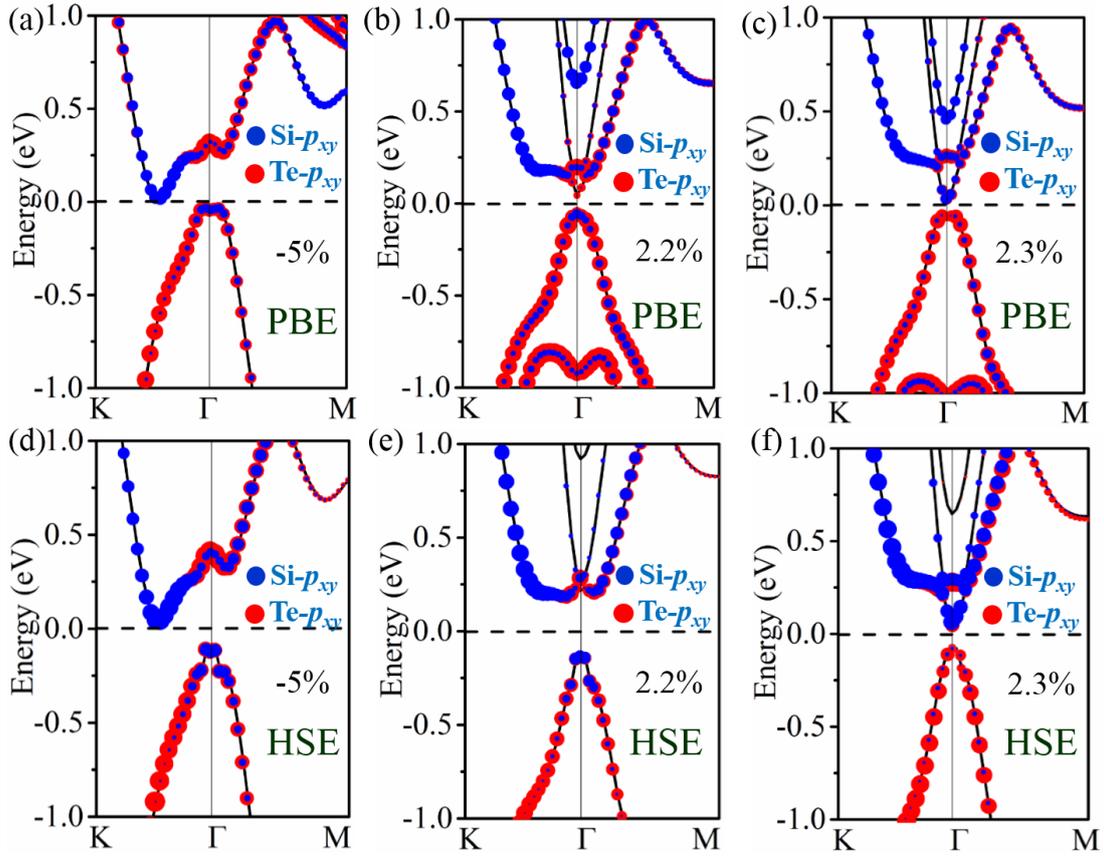

**Fig. S2** Orbital-resolved band structures with SOC of $Si_2Te_2$ by PBE method under the value of strain ε= -5.0% (a), 2.2% (b), 2.3% (c), respectively. (d)-(f) present orbital-resolved band structures with SOC via HSE method under strains ε = -5.0%, 2.2%, 2.3% respectively. The red dots represent the contributions from the Te-$p_{x,y}$ orbitals and the blue dots represent contributions from the $p_{x,y}$ orbitals of Si atom.